\documentclass[
               ]{jacow}
\usepackage{pdfpages,multirow,ragged2e} %
\makeatletter%
	\ifboolexpr{bool{xetex}}
	 {\renewcommand{\Gin@extensions}{.pdf,%
	                    .png,.jpg,.bmp,.pict,.tif,.psd,.mac,.sga,.tga,.gif,%
	                    .eps,.ps,%
	                    }}{}
\makeatother

\ifboolexpr{bool{xetex} or bool{luatex}} 
 {}                                      
 {\usepackage[utf8]{inputenc}}           

\usepackage[english]{babel}
\usepackage{siunitx}
\begin{document}

\title{STATUS OF ACCELERATOR AUTOMATION IN THE COMMUNITY: CLASSIFICATION, SURVEY RESULTS, AND THE ROLE OF AI AND MACHINE LEARNING}

\author{S.~Reimann\thanks{s.reimann@gsi.de}\footnote{also at  Facility for Antiproton and Ion Research, Darmstadt, Germany}, GSI Helmholtz Centre for Heavy Ion Research, Darmstadt, Germany\\ 
F.~Miyahara, High Energy Accelerator Research Organization, Tsukuba, Japan\\ 
M.~Lafky, Australian Synchrotron, Melbourne, Australia\\ 
P.~Schuh, SLAC National Accelerator Laboratory, Menlo Park, United States\\ 
J.~Xing, Institute of High Energy Physics, Beijing, China\\ 
G.~Johns, Oak Ridge National Laboratory, Oak Ridge, United States\\ 
R.~Steerenberg, European Organization for Nuclear Research, Geneva, Switzerland\\ 
B.~Freeman, Thomas Jefferson National Accelerator Facility, Newport News, United States\\ 
K.~Genge, TRIUMF, Vancouver, Canada\\ 
M.~Pont, ALBA Synchrotron (Spain), Cerdanyola del Vallès, Spain\\ 
G.~Marr, Brookhaven National Laboratory, Upton, United States\\ 
Y.~Iwata, National Institutes for Quantum Science and Technology, Chiba, Japan\\ 
L.~Hardy, European Synchrotron Radiation Facility, Grenoble, France}

\maketitle


%
\begin{abstract}
The increasing complexity of particle accelerators and the growing demand for operational efficiency 
have intensified the focus on automation in accelerator control systems. However, there is still a lack 
of a common language to describe and compare the level of automation across different facilities. 
This article presents a generic classification scheme for accelerator automation, defining five distinct 
levels ranging from basic operational assistance to full autonomy. To assess the current state of 
automation in the accelerator community, a structured survey was conducted during the 
14th Workshop on Accelerator Operations (WAO~2025). Responses from major laboratories indicate that more than one third of accelerators 
currently operate at the lowest level~1 (operational assistance), and no facility has yet reached full 
automation. However, projections show a clear trend toward higher levels within the next decade. 
Key drivers include optimization algorithms for autotuning, finite-state machines, and closed-loop 
feedback systems. This study provides both a conceptual framework and a snapshot of current 
implementation practices, supporting a shared understanding of accelerator automation across 
the community.
\end{abstract}

\section{Levels of Accelerator Automation}

The classification introduced in this work is conceptually inspired by SAE J3016~\cite{SAEJ3016}, 
a widely adopted five-level framework for driving automation. Although originally developed for on-road motor vehicles, its structured progression—from full manual control to full autonomy—has proven useful as a generic reference for describing automation maturity. When transferring this concept to accelerator operations, it is adapted to the specific system structure and operational logic of accelerator facilities. Based on this adaptation, we propose the following definition of accelerator automation levels.

\subsection{Level 0: No Automation}
At this level, all control actions are performed manually.
Operators directly adjust technical parameters such as currents and voltages without the support of automated systems.

\subsection{Level 1: Operational Assistance}
Initial assistance systems become available to support accelerator operation. These include parameter hierarchy models, beam loss and transmission monitoring, and software-based interlock mechanisms. Importantly, tuning is performed using high-level physics parameters rather than raw hardware settings, thereby reducing operational complexity.

\subsection{Level 2: Partial Automation}
This level introduces true automation into the operational workflow. Parameter-scanning tools and finite-state-machine-based sequencers are used to automate repetitive tasks. In addition, autotuning solutions are available, allowing operators to initiate optimization routines that adjust multiple set values to achieve defined machine objectives.

\subsection{Level 3: Conditional Automation}
The accelerator can operate autonomously under predefined and stable conditions.
Once initial configuration is complete, performance is maintained through automatic feedback loops, stabilization routines, and self-correction mechanisms.
Operators are required only when machine goals or boundary conditions change.

\subsection{Level 4: High-Level Automation}
All major operational procedures—such as re-commissioning, RF/HV conditioning, machine setup, stabilization, and fine-tuning—are fully automated.
Operators are mainly responsible for initiating and supervising these processes, rather than executing them manually.
Human oversight remains essential, but the system is capable of performing complex sequences independently.

\subsection{Level 5: Full Automation}
At this highest level, no human operator is required for routine operation.
The accelerator system is fully autonomous in operational control, handling startup, configuration, optimization, controlled shutdown, and recovery from operational faults within the technical design envelope without operator intervention.
Faults requiring physical repair, restoration of infrastructure or communication, or other on-site technical intervention are outside the scope of automation; in such cases, the system is expected to reach or maintain a safe state and automatically alert the appropriate expert personnel.
Once the required technical intervention has restored the necessary boundary conditions, restart and recommissioning can proceed autonomously.
While this level remains aspirational for accelerator facilities, it represents the ultimate vision of automation.

\section{Survey Methodology}

To assess the current status and future trajectory of accelerator automation, 
an online survey was conducted in conjunction with the 14th Workshop on Accelerator Operations (WAO~2025), 
held in September 2025 in Saskatoon, Canada~\cite{WAO2025}. The survey was developed by the authors and made available to workshop participants via an online form.

The questionnaire collected basic metadata including the participant's name, institution, accelerator name, 
and accelerator type. The core questions were grouped into three categories: automation level, time horizon, 
and main automation drivers. Based on the five-level classification scheme introduced above, each respondent 
was asked to indicate the automation level:
\begin{itemize}[noitemsep, topsep=3pt]
  \item at the time of the survey (Sept. 2025),
  \item after implementation of already planned upgrades,
  \item and as predicted in approximately ten years.
\end{itemize}

In addition, respondents were asked to name the main driver(s) for advancing automation at their facility, 
selected from an open list (e.g., FSMs, autotuning, machine learning, feedback systems). 
An optional free-text field allowed for further clarification or comments.

A total of \textbf{29 complete responses} were received and used as the basis for the analysis.
For two accelerators, independent assessments were provided by two individuals each.
In four out of six comparisons, the reported automation levels differed by one level.
To estimate the variability between assessments, the reported levels in these cases were mapped onto a locally normalized two-level scale.
On this scale, the mean value $\bar{l} = 2/3$ was obtained, which was used solely for estimating the standard deviation:
\[
s = \sqrt{ \frac{1}{n - 1} \sum (l_i - \bar{l})^2 }
  \approx 0.5 .
\]
The standard deviation $s$ quantifies the typical disagreement between independent automation level assessments and indicates that classifications differ by roughly half a level.

\section{Results}

Responses from major laboratories worldwide were received, representing a broad spectrum of accelerator facilities. The participating institutes include ANL, BNL~\cite{Niedziela:ICALEPCS07-ROPB01, Lin:hiat2025-tux03, DOttavio:ICALEPCS01-THAP062}, CLS, CERN~\cite{Foldesi:ipac2025-thpm009}, Diamond, ESS, Fedoruk Centre, GSI/FAIR~\cite{Appel:IJMPA2019, Appel:IPAC19-WEPMP005, Geithner:IPAC18-THPML028, Reimann:IPAC22-TUPOST031, Reimann:IPAC25-THBN2}, ISIS~\cite{Leputa:ICALEPCS25-WEPD078, Saoulis:IPAC22-TUPOPT057}, JLab~\cite{Colen:MLST2026}, KEK~\cite{Natsui:IPAC25-WEXN1}, Los Alamos National Laboratory, MAX IV, MedAustron, ORNL, Rutherford Appleton Laboratory, SLAC~\cite{Edelen2026, Mishra:arXiv2505.23858}, and TRIUMF~\cite{Ghelfi:RSI2025}.

The survey covered a wide range of accelerator types: 
LINACs (\SI{34.5}{\percent}), Synchrotron Light Sources (\SI{17.2}{\percent}), 
other Synchrotrons (\SI{20.7}{\percent}), Colliders (\SI{10.3}{\percent}), 
Cyclotrons (\SI{10.3}{\percent}), one Storage Ring (\SI{3.4}{\percent}), and one
Recirculating LINAC (\SI{3.4}{\percent}).

\begin{figure}[!htb]
   \centering
   \includegraphics[width=1.0\columnwidth]{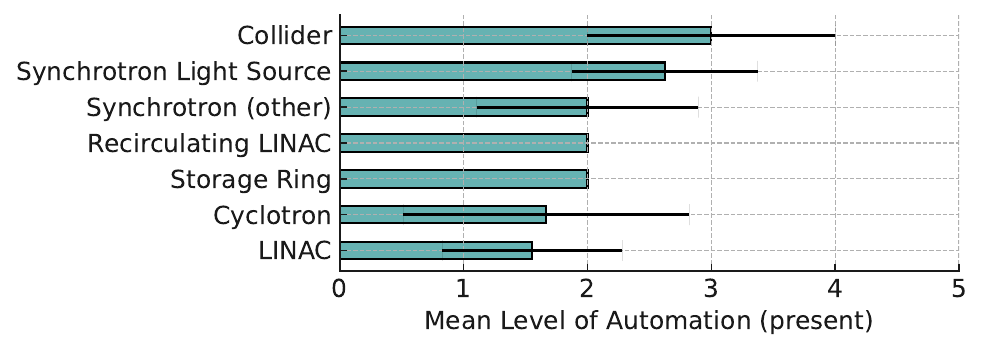}
	\caption{Mean level of automation for different accelerator types based on the current operational state (\textit{present level of automation}). Error bars indicate the standard deviation of the responses within each category.}

   \label{fig:automation_by_type_present}
\end{figure}

The present level of automation across the various accelerator types is illustrated in Fig.~\ref{fig:automation_by_type_present}. 
The results suggest that colliders and synchrotron light sources exhibit, as anticipated, a relatively higher degree of operational automation. In contrast, cyclotrons and linear accelerators remain at the lower end of the automation scale.

\begin{figure}[!htb]
   \centering
   \includegraphics[width=1.0\columnwidth]{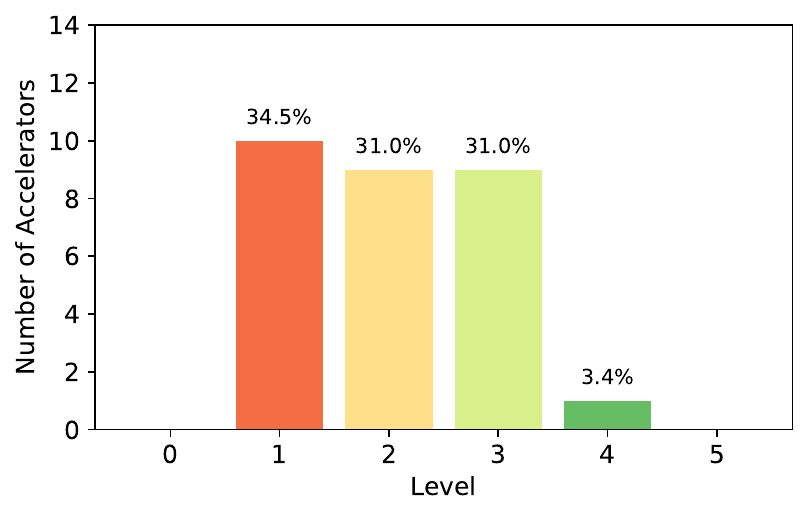}
   \caption{Present level of automation.}
   \label{fig:present_level}
\end{figure}

\begin{figure}[!htb]
   \centering
   \includegraphics[width=1.0\columnwidth]{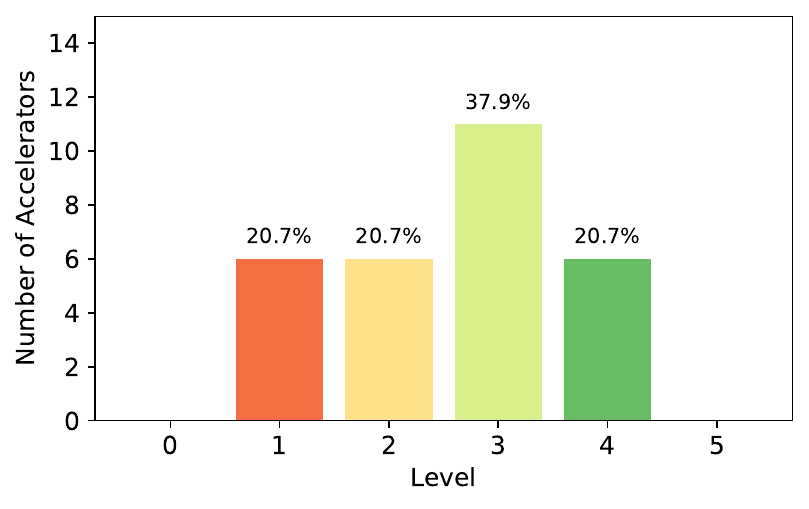}
   \caption{Planned level of automation.}
   \label{fig:planned_level}
\end{figure}

\begin{figure}[!htb]
   \centering
   \includegraphics[width=1.0\columnwidth]{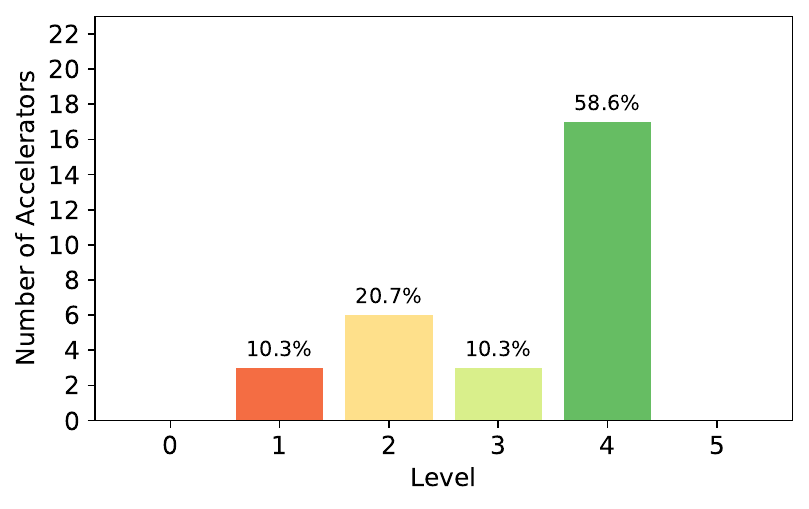}
   \caption{Predicted level of automation in 2035.}
   \label{fig:future_level}
\end{figure}

The following figures illustrate the distribution of automation levels across all survey responses. 
Figure~\ref{fig:present_level} presents the current status as of September~2025, 
while Fig.~\ref{fig:planned_level} reflects the expected state after completion of all already approved and planned upgrade projects. 
In addition, all participants were asked to provide an estimate of how the level of automation is likely to evolve over the next decade. 
This projection is summarized in Fig.~\ref{fig:future_level}.


According to the submitted data, manual-only operation has essentially disappeared from the accelerator community. 
However, full automation has not yet been implemented at any facility. 
At present, approximately one third of all accelerators — corresponding to Level~1 — are still operated without genuine automation solutions. 
Level~4 automation has so far only been reported for the LHC.

A clear trend toward higher automation levels can nevertheless be observed. 
Once all currently approved and planned measures are implemented, nearly 60\,\% of the facilities are expected to reach at least the level of \textit{Conditional Automation} (Level~3). 
The long-term projection indicates an almost 60\,\% share at Level~4. 
While these forecasts appear optimistic overall, around one in ten respondents anticipate that their accelerators will still operate without true automation solutions even ten years from now. 
Notably, none of the participating laboratories considered full autonomy (Level~5) to be a realistic target within this time frame.

\section{Enabling Technologies}

\begin{figure}[!htb]
   \centering
   \includegraphics[width=1.0\columnwidth]{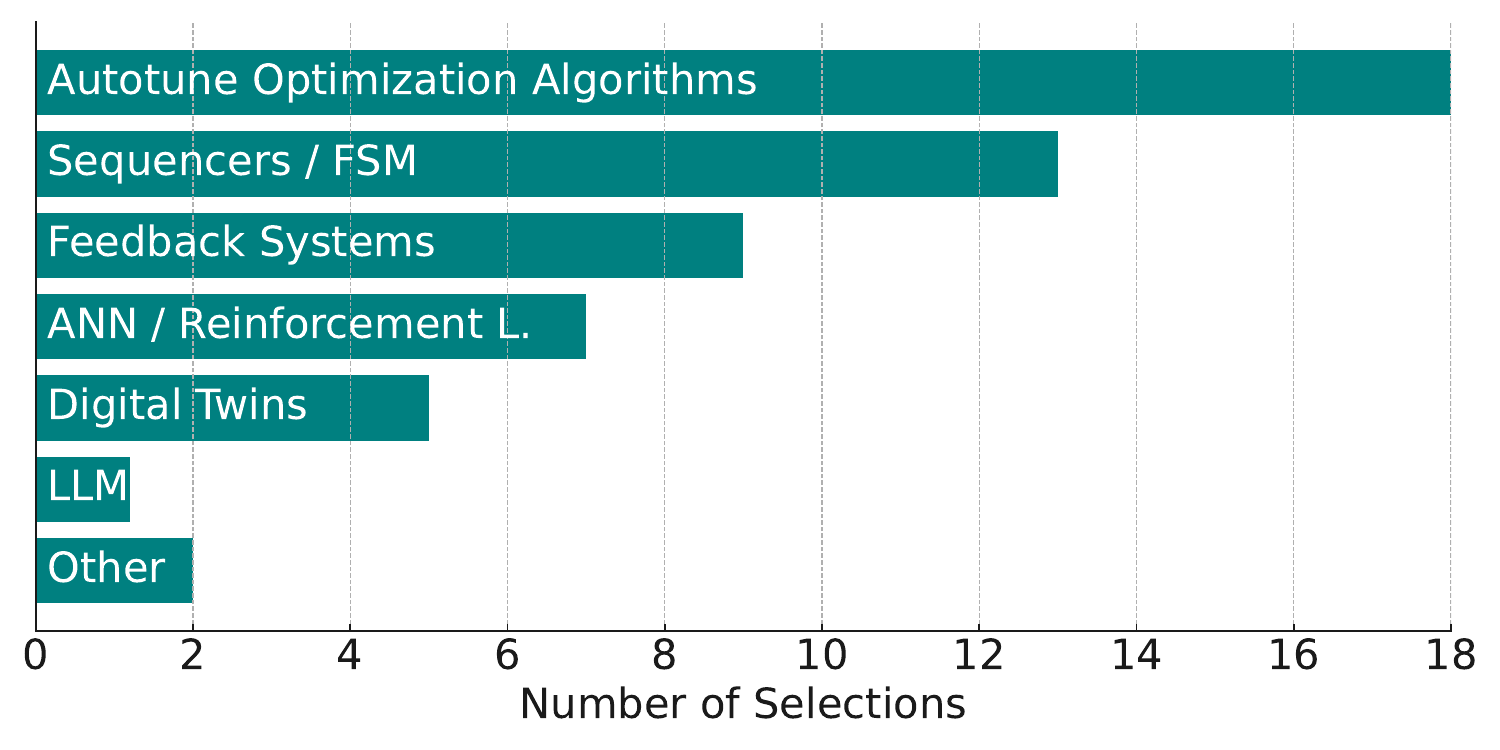}
   \caption{Expected drivers of the next step in accelerator automation. 
Multiple selections were allowed. The most frequently selected driver category was \emph{Autotune Optimization Algorithms}, 
including examples such as BOBYQA, Bayesian optimization, and Genetic algorithms. 
FSM = Finite-State Machines, ANN = Artificial Neural Networks, LLM = Large Language Models.}
   \label{fig:technology}
\end{figure}

Figure~\ref{fig:technology} summarizes the technologies identified by the participants as the most relevant drivers of automation. 
By a considerable margin, auto-tuning algorithms (mentioned by 62\,\% of respondents) and sequencer systems (45\,\%) were rated as the most important enablers. 
They are followed by feedback systems (31\,\%). 
Artificial neural networks, digital twins (and simulation tools) were mentioned less frequently, while large language models (LLMs) and transformers currently remain a marginal topic within the accelerator community — 
despite their recent popularity and prominence in the broader field of artificial intelligence.

\section{Conclusion and Outlook}

A structured taxonomy for classifying automation levels in accelerator operation has been introduced, providing a common framework for assessing the degree of automation across different facilities. 
The survey results show a relatively even distribution among Levels~1 to~3, whereas Levels~0 and~5 do not occur, even in the long-term projections. 
Among the enabling technologies, auto-tuning algorithms and finite state machines are currently regarded as the most promising approaches, followed by feedback systems. ANNs and LLMs play only a minor role.
A broader dataset with contributions from additional laboratories, as well as periodic re-evaluations, would be valuable to monitor the evolution of automation practices over time.

\section{ACKNOWLEDGEMENTS}

The authors would like to thank the local organizers and participants of the WAO~2025 workshop for their valuable contributions to this study.

%
%

%
%


\end{document}